\documentclass[prc,aps,showkeys,superscriptaddress,nofootinbib,floatfix]{revtex4}
\usepackage{graphicx} 

\begin{document}

\title{Glow in a stream of charged particles}

\author{B.V. Voitsekhovskii}
\affiliation{Hydrodynamics Institute, Novosibirsk, Russia}
\author{B.B. Voitsekhovskii}
\affiliation{Hydrodynamics Institute, Novosibirsk, Russia}
\date{January 5, 1976}

\begin{abstract}
The results of an experiment with a generator of a stream of charged drops are reported. The glow of subjects placed in the stream is observed. 
The volume of the glowing region reaches 20 cm$^3$ at a current less than 20 $\mu$A through the object. Ideas are expressed concerning the connection between St.Elmo's fire and the observed glow.
\end{abstract}
\keywords{St.Elmo's fire}
\maketitle
 
 Most phenomena connected with atmospheric electricity occur in clouds of charged particles.
 When atmospheric-electricity phenomena are simulated in the laboratory, it is difficult to produce clouds commensurate 
 with natural ones, so that to duplicate natural effects the charge density of the laboratory clouds must be much higher.
 In a thunderstorm cloud, where the density of the excess charge is maximal, it reaches 100 cgs esu/m$^3$ in
 regions having a linear dimension 50 m~\cite{cloud}.
 Our interest in the production of charged-particle clouds is connected, in particular, with the theory 
 proposed in~\cite{ball} for ball lightning, the production of which calls for a high density of the excess charge.
 
 1. We have developed a generator that produces clouds of charged drops.
 It emits a continuous jet of air (air flow $\sim$ 10 liter/sec), and the jet carries $\sim$ 5 cm$^3$/sec
 of liquid in the form of minute drops of $\sim 10 \, \mu$m diameter.
 The drops carry up to 30 micro Coulomb per second.
 The droplet current depends little on the flow of the liquid and increases linearly with increasing air flow.
 After leaving the generator, the jet is rapidly slowed down by friction and by mixing with the surrounding air.
 This produces strong electric fields.
 The construction of large generators with current $\sim$ 3 mA will apparently make it possible to obtain within 
 10-100 sec clouds in which lightning discharges close to natural can be produced.
 A detailed description of the generator and of experiments aimed at increasing the droplet current will be 
 presented in a supplementary communication.\\
 
 2. The study of the stream of charged water droplets has revealed that various objects placed in the stream are made to glow.
 Figure 1 shows a photograph of the glow of the fingers of a hand and of a steel cone.
 The current flowing off the object and the dimension of the glowing region decrease with increasing distance 
 from the object to the generator, but the glow is noticeable even at a distance of 3 m.
 When the object is taken out of the stream in a radial direction, the glow first increases and is maximal at a distance 
 2-5 cm from the stream boundary on the outside of the stream.\\
 
 The current for the glowing objects shown in the figure did not exceed 20 $\mu$A.
 The glow of sharply pointed objects had the shape of a fan diverging from the pointed end in the direction towards the stream.
 If a thin rod (pencil) is placed longitudinally in the stream, the glow is produced on both its ends.\\
 
 Screening is observed, i.e., if a second object is placed (closer to the generator), then the glow of the far object becomes weaker.
 In one of the experiments we observed the glow of a fishing line (Fig. 2) whose ends were outside the stream and 
 remained dry on the outside.
 To improve the insulation, we made an assembled ebonite rod.
 Immediately after removal from a water bath, its resistance was $\sim 10^{12} \, \Omega$.
 We were unable to observe any glow from short objects placed in the stream with the aid of the rod.
However, when a long object was mounted on the rod, a weak glow was observed, apparently due to leakage 
to the air outside the stream.\\
 
 Experiments were set up with two streams of air with charged water drops of opposite polarity.
 The generators were placed opposite each other in such a way that the streams came in contact, 
 and a glow could be expected in the contact region. 
 However, neither in this arrangement nor with parallel streams was glow observed in the region where the streams came in contact.
 This result means that charged particles of pure water are not sufficient to produce glow.
 The possible reason is that the energy required to produce an ion drawn out of the water is high.
 In this case a water jet aimed into a stream of charged particles should not glow, as was indeed confirmed by experiment.
 On the other hand if a jet of emulsion is directed into the stream, its end glows brightly (Fig. 3).\\
 
 It is of interest to compare the phenomena observed by us with Saint Elmo's fire.
 Saint Elmo's fire is produced in bad weather on the tops of masts, on mountains with sharp rocks, and on trees and similar objects.
 Many authors~\cite{lightning,atmospheric} regard Saint Elmo's fire as a corona-discharge glow.
 However, at the same current flowing through an object, the glow due to corona is weaker than the glow in a stream of charged drops.
 Poor weather is frequently accompanied by a horizontal flow of air carrying low-lying clouds. 
 The lower parts of the clouds are negatively charged and tall objects enter the horizontal stream of negatively charged water drops.
 These are exactly the conditions duplicated in our experiment.
 All this allows us to assume that Saint Elmo's fire and the glow observed by us in a stream of charged drops are of the same nature.\\
 
 CONCLUSIONS\\
 
 1. We have constructed a generator for a stream of charged drops with a current up to 30 $\mu$A and average drop radius $\sim$10 $\mu$.\\
 
 2. We observed the glow of various objects placed in the stream, with appreciable brightness, at a glow-region volume $\sim$ cm$^3$
 and a current less than 20  $\mu$A through the object.\\
 
 3. The nature of the well known Saint Elmo fire is apparently analogous to that of the glow observed by us.
 A stream of charged particles produces glow around the object both in nature and in our experiment.
 
\begin{figure}
\includegraphics[trim = 0mm 0mm 0mm 0mm, width=1\linewidth]{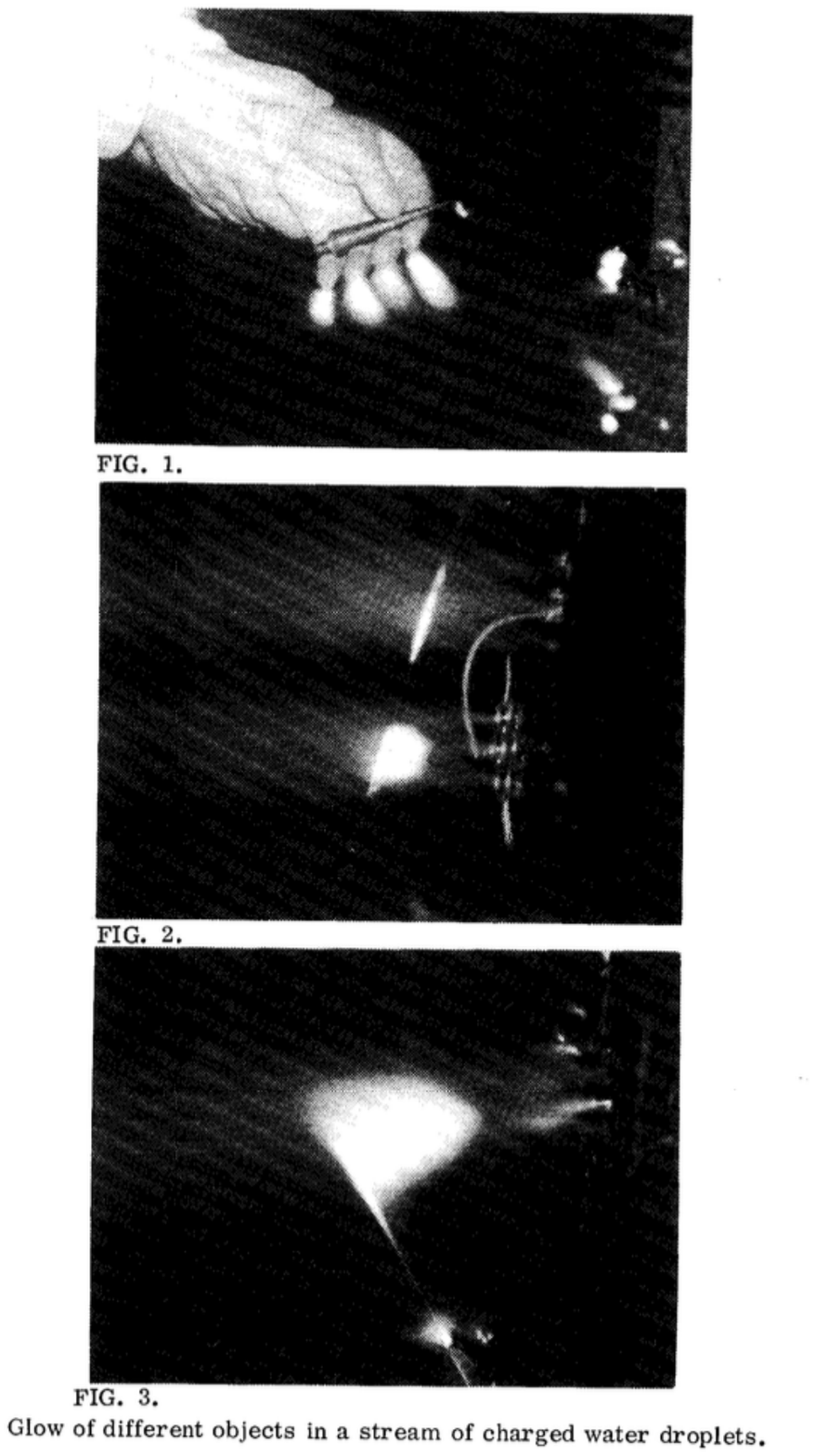}
\end{figure}

\end{document}